\pdfoutput=1
\documentclass[12pt]{article}

\usepackage{epsfig}

\usepackage{times}

\newcounter{lastnote}

\title{\Large Radio Imaging of the Very-High-Energy $\gamma$-Ray
Emission Region in the Central Engine of a Radio Galaxy}

\author{
The VERITAS Collaboration, the VLBA 43\,GHz M\,87 Monitoring Team,\\
the H.E.S.S. Collaboration, and the MAGIC Collaboration\footnote{To whom
correspondence should be addressed; E-mail: beilicke@physics.wustl.edu,
or krawcz@wuphys.wustl.edu (VERITAS), cwalker@aoc.nrao.edu, or
phardee@bama.ua.edu (VLBA), martin.raue@mpi-hd.mpg.de (H.E.S.S.),
mazin@ifae.es, or robert.wagner@mpp.mpg.de (MAGIC).}
\\
\footnotesize{The full author list with affiliations can be found at the 
end of this paper}
}

\date{}

\begin{document} 



\maketitle 

\begin{abstract}

The accretion of matter onto a massive black hole is believed to feed the
relativistic plasma jets found in many active galactic nuclei (AGN).
Although some AGN accelerate particles to energies exceeding
$10^{12}$ electron Volts (eV) and are bright sources of very-high-energy
(VHE) $\gamma$-ray emission, it is not yet known where the VHE emission
originates. Here we report on radio and VHE observations of the radio
galaxy M\,87, revealing a period of extremely strong VHE $\gamma$-ray
flares accompanied by a strong increase of the radio flux from its
nucleus. These results imply that charged particles are accelerated to
very high energies in the immediate vicinity of the black hole.

\end{abstract}



Active galactic nuclei (AGN) are extragalactic objects thought to be
powered by massive black holes in their centres. They can show strong
emission from the core, which is often dominated by broadband continuum
radiation ranging from radio to X-rays and by substantial flux
variability on different time scales. More than 20 AGN have been
established as VHE $\gamma$-ray emitters with measured energies above
$0.1$ tera electron Volts (TeV); the jets of most of these sources are
believed to be aligned with the line-of-sight to within a few degrees.
The size of the VHE $\gamma$-ray emission region can generally be
constrained by the time scale of the observed flux variability
\cite{Mrk421_Burst,M87_HESS} but its location remains unknown.

\begin{figure}[p]

\centering{
\epsfig{file=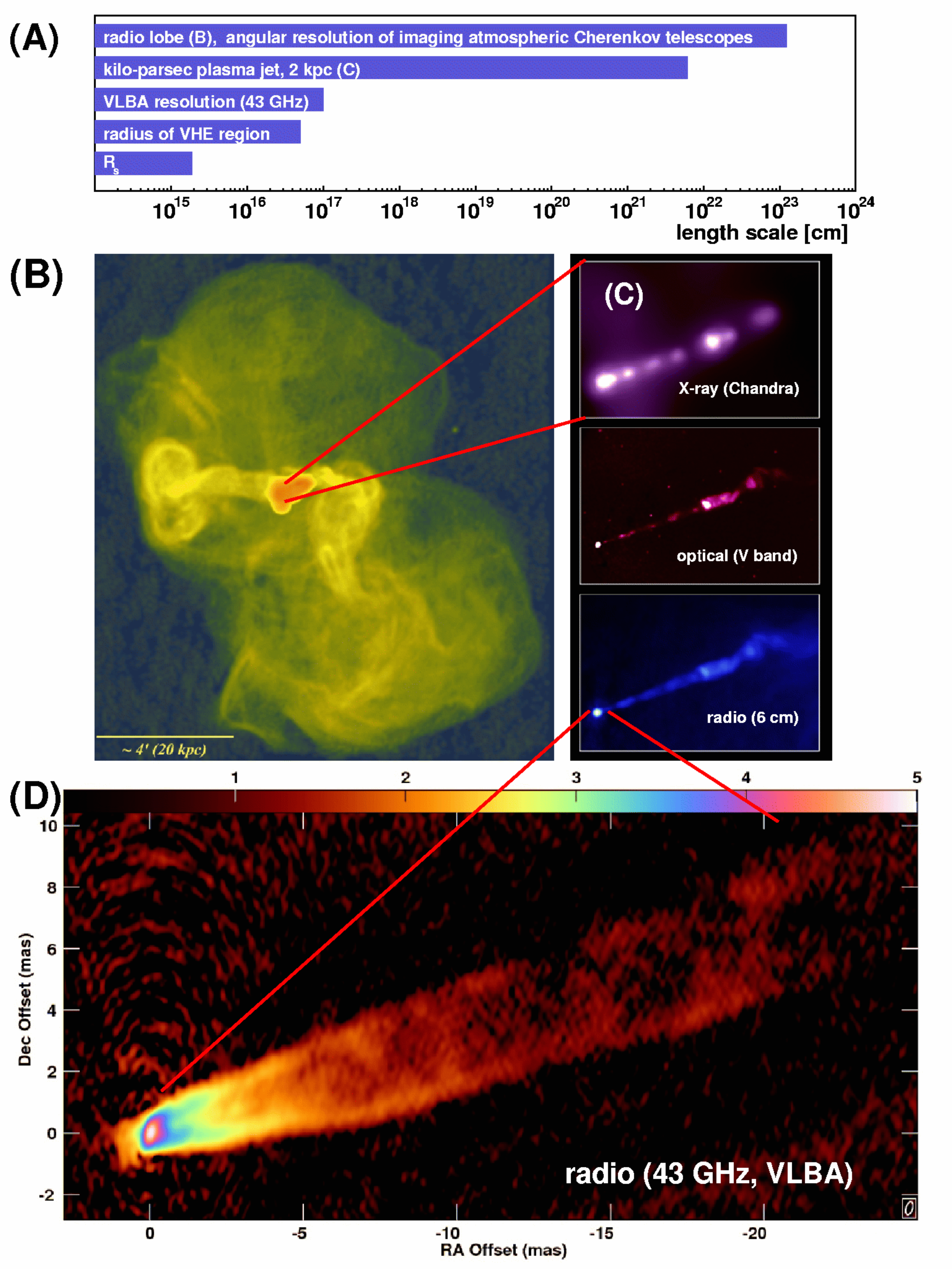,width=0.75\textwidth}
}

\caption{\label{fig1} \footnotesize M\,87 at different photon frequencies
and length scales.  ({\bf A}): Comparison of the different length scales.
({\bf B}): $90 \, \rm{cm}$ radio emission measured with the VLA. The jet
outflows terminate in a halo which has a diameter of roughly $80 \,
\rm{kpc}$ ($15'$). The radio emission in the central region is saturated
in this image. Credit: F.N.~Owen, J.A.~Eilek and N.E.~Kassim
\cite{M87_90CM}, NRAO/AUI/NSF. ({\bf C}): Zoomed image of the plasma jet
with an extension of $2 \, \rm{kpc}$ ($20''$), seen in different frequency
bands: X-rays (Chandra, upper panel) optical (V band, middle) and radio
($6 \, \rm{cm}$, lower panel). Individual knots in the jet and the nucleus
can be seen in all three frequency bands. The innermost knot HST-1 is
located at a projected distance of $0.86 \, \rm{arcseconds}$ ($60 \,
\rm{pc}$, $\approx 10^{5} \, R_{\rm{s}}$) from the nucleus. Credit: X-ray:
NASA/CXC/MIT/H.~Marshall et al., radio: F.~Zhou, F.~Owen (NRAO),
J.~Biretta (STScI), optical: NASA/STScI/UMBC/E.~Perlman et al.,
\cite{ChandraSpecM87}. ({\bf D}): An averaged, and hence smoothed, radio
image based on 23 images from the VLBA monitoring project at $43 \,
\rm{GHz}$. The color scale gives the logarithm of the flux density in
units of $0.01 \, \rm{mJy/beam}$. The indication of a counter-jet can be
seen, emerging from the core towards the lower left side. Image from
\cite{JointPaper}.}

\end{figure}

We studied the inner structure of the jet of the giant radio galaxy M\,87,
a known VHE $\gamma$-ray emitting AGN
\cite{M87_HEGRA,M87_HESS,M87_VERITAS,M87_MAGIC} with a $(6.0 \pm 0.5)
\times 10^{9} \, \rm{M}_{\odot}$ black hole \cite{BH_M87}, scaled by
distance, located $16.7 \, \rm{Mpc}$ (54 million light years) away in the
Virgo cluster of galaxies. The angle between its plasma jet and the
line-of-sight is estimated to lie between $15-25 \deg$ (see supporting
online text). The substructures of the jet, which are expected to scale
with the Schwarzschild radius $R_{\rm{s}}$ of the black hole\footnote{The
Schwarzschild radius of a black hole with the mass $m$ is defined as
$R_{\rm{s}} = 2 G m / c^{2}$, $G$ is the gravitational constant, and $c$
is the speed of light. The Schwarzschild radius defines the event horizon
of the black hole.}, are resolved in the X-ray, optical and radio
wavebands \cite{ChandraSpecM87} (Fig.~\ref{fig1}). High-frequency radio
very long baseline interferometry (VLBI) observations with
sub-milliarcsecond (mas) resolution are starting to probe the collimation
region of the jet \cite{JetFormation}. With its proximity, bright and
well-resolved jet, and very massive black hole, M\,87 provides a unique
laboratory in which to study relativistic jet physics in connection with
the mechanisms of VHE $\gamma$-ray emission in AGN.

VLBI observations of the M\,87 inner jet show a well resolved,
edge-brightened structure extending to within $0.5 \, \rm{mas}$ ($0.04 \,
\rm{pc}$ or $70 \, R_{\rm{s}}$) of the core. Closer to the core, the jet
has a wide opening angle suggesting that this is the collimation region
\cite{JetFormation}. Generally, the core can be offset from the actual
location of the black hole by an unknown amount \cite{Marscher}, in which
case it could mark the location of a shock structure or the region where
the jet becomes optically thin. However, in the case of M\,87 a weak
structure is seen on the opposite side of the core from the main jet,
which may be the counter-jet, based on its morphology and length
\cite{VLBA_M87,InnerJet}. Together with the observed pattern in opening
angles, this suggests that the black hole of M\,87 is located within the
central resolution element of the VLBI images, at most a few tens of
$R_{\rm{s}}$ from the radio core (see supporting online text). Along the
jet, previous monitoring observations show both near-stationary components
\cite{InnerJet} (pc-scale) and features that move at apparent superluminal
speeds \cite{HST_Superluminal,Jet_And_TeV} (100 pc-scale). The presence of
superluminal motions and the strong asymmetry of the jet brightness
indicate that the jet flow is relativistic. The near-stationary components
could be related to shocks or instabilities, that can be either
stationary, for example if they are the result of interaction with the
external medium, or slowly moving if they are the result of instabilities
in the flow.

A first indication of VHE $\gamma$-ray emission from M\,87 was reported by
the High Energy Gamma-Ray Astronomy (HEGRA) collaboration in 1998/99
\cite{M87_HEGRA}. The emission was confirmed by the High Energy
Stereoscopic System (H.E.S.S.) in 2003-2006 \cite{M87_HESS}, with
$\gamma$-ray flux variability on time scales of days. M\,87 was detected
again with the Very Energetic Radiation Imaging Telescope Array System
(VERITAS) in 2007 \cite{M87_VERITAS} and, recently, the short-term
variability was confirmed with the Major Atmospheric Gamma-Ray Imaging
Cherenkov (MAGIC) telescope during a strong VHE $\gamma$-ray outburst
\cite{M87_MAGIC} in February 2008. Causality arguments imply that the
emission region should have a spatial extent of less than $\approx 5
\delta R_{\rm{s}}$, where $\delta$ is the relativistic Doppler factor.
This rules out explanations for the VHE $\gamma$-ray emission on the basis
of (i) dark matter annihilation \cite{NeutralinoM87}, (ii) cosmic-ray
interactions with the matter in M\,87 \cite{CR_M87}, or (iii) the knots in
the plasma jet (Fig.~\ref{fig1}C). Leptonic \cite{ModelGeorg,ModelLenain}
and hadronic \cite{ModelReimer} VHE $\gamma$-ray jet emission models have
been proposed. However, the location of the emission region is still
unknown. The nucleus \cite{ModelNeronov,ModelRieger}, the inner jet
\cite{ModelTav} or larger structures in the jet, such as the knot HST-1
(Fig.~\ref{fig1}C), have been discussed as possible sites
\cite{Jet_And_TeV}. Because the angular resolution of VHE experiments is
of the order of $0.1 \deg$, the key to identifying the location of the VHE
$\gamma$-ray emission lies in connecting it to measurements at other
wavebands with considerably higher spatial resolutions. An angular
resolution more than six orders of magnitude better (less than $6 \times
10^{-8}$ degrees, corresponding to approximately $30 \, R_{\rm{s}}$ in
case of M\,87) can be achieved with radio observations (Fig.~\ref{fig1}).

\begin{figure}[p]

\centering{
\epsfig{file=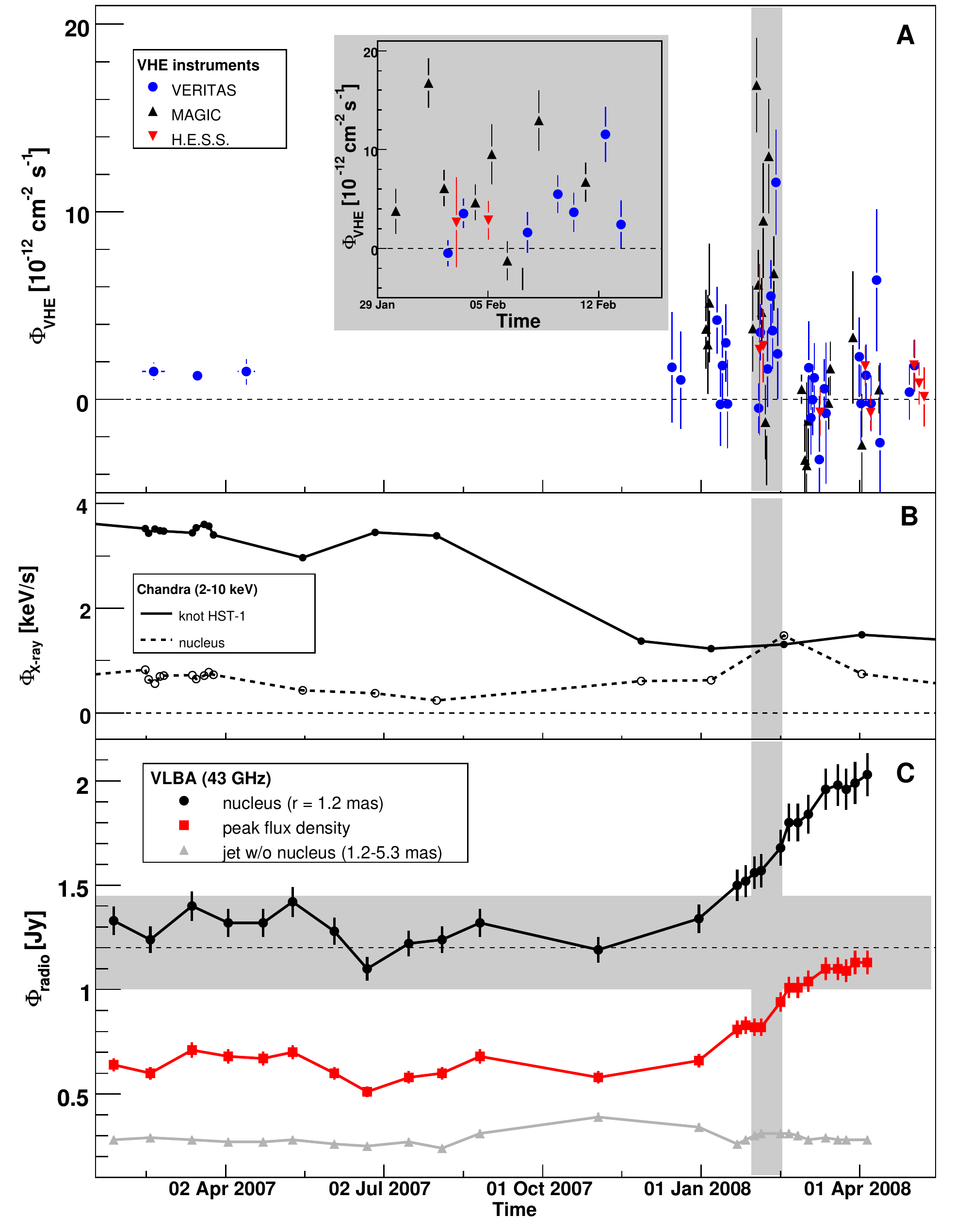,width=0.75\textwidth}
}

\caption{\label{fig2} \footnotesize Combined M\,87 light curves from 2007
to 2008. ({\bf A}): VHE $\gamma$-ray fluxes ($E > 0.35 \, \rm{TeV}$,
nightly average), showing the H.E.S.S., MAGIC and VERITAS data. The fluxes
with statistical errors (1 standard deviation) were calculated assuming a
power-law spectral shape of $\rm{d}N / \rm{d}E \propto E^{-2.3}$.
Monthly-binned archival VERITAS data taken in 2007 are also shown
\cite{M87_VERITAS}. The systematic uncertainty in the flux calibration
between the experiments was estimated to be on the order of $20 \%$ based
on Crab Nebula data. The regular gaps in the light curve correspond to
phases of full moon during which no observations were possible. The inlay
shows a zoomed version of the flaring activity in February 2008; the time
span is indicated by the grey vertical box in all panels. ({\bf B}):
Chandra X-ray measurements ($2 - 10 \, \rm{keV}$) of the nucleus and the
knot HST-1 \cite{TimeScalesHST}. ({\bf C}): Flux densities from the $43 \,
\rm{GHz}$ VLBA observations are shown for (i) the nucleus (circular region
with radius $r = 1.2 \, \rm{mas} = 170 \, R_{\rm{s}}$ centered on the peak
flux), (ii) the peak flux (VLBA resolution element), and (iii) the flux
integrated along the jet between distances of $r = 1.2 - 5.3 \, \rm{mas}$
(compare with Fig.~\ref{fig3}). The error bars correspond to $5 \%$ of the
flux. The shaded horizontal area indicates the range of fluxes from the
nucleus before the 2008 flare. Whereas the flux of the outer regions of
the jet does not change substantially, most of the flux increase results
from the region around the nucleus. Image from \cite{JointPaper}.}

\end{figure}

We used the H.E.S.S. \cite{Crab_HESS}, MAGIC \cite{Crab_MAGIC} and VERITAS
\cite{VERITAS_LSI} instruments to observe M\,87 during 50 nights between
January and May 2008, accumulating over $95 \, \rm{h}$ of data (corrected
for the detector dead times) in the energy range between $0.1 \, \rm{TeV}$
and several 10's of TeV. Simultaneously, we monitored M\,87 with the Very
Long Baseline Array \cite{VLBA} (VLBA) at $43 \, \rm{GHz}$ with a
resolution of $0.21 \times 0.43 \, \rm{mas}$ \cite{ImagingJetBase},
corresponding to about $30 \times 60 \, R_{\rm{s}}$, see
\cite{JointPaper}. During the first half of 2008, three X-ray pointings
were performed with the Chandra satellite \cite{TimeScalesHST}. Our light
curves are shown in Fig.~\ref{fig2}.

We detected multiple flares at VHE in February 2008 with denser sampling,
following a trigger sent by MAGIC [$\sim$23~h of the data published in
\cite{M87_MAGIC}]. The short-term VHE variability, first observed in 2005
\cite{M87_HESS}, is clearly confirmed and the flux reached the highest
level observed so far from M\,87, amounting to more than $10 \%$ of that
of the Crab Nebula. At X-ray frequencies the innermost knot in the jet
(HST-1) is found in a low state, whereas in mid February 2008 the nucleus
was found in its highest X-ray flux state since 2000 \cite{TimeScalesHST}.
This is in contrast to the 2005 VHE $\gamma$-ray flares \cite{M87_HESS},
which happened after an increase of the X-ray flux of HST-1 over several
years \cite{HST1_Burst}, allowing speculation that HST-1 might be the
source of the VHE $\gamma$-ray emission \cite{Jet_And_TeV}; no $43 \,
\rm{GHz}$ radio observations were obtained at that time. Given its low
X-ray flux in 2008, HST-1 is an unlikely site of the 2008 VHE flaring
activity.

\begin{figure}[p]

\centering{
\epsfig{file=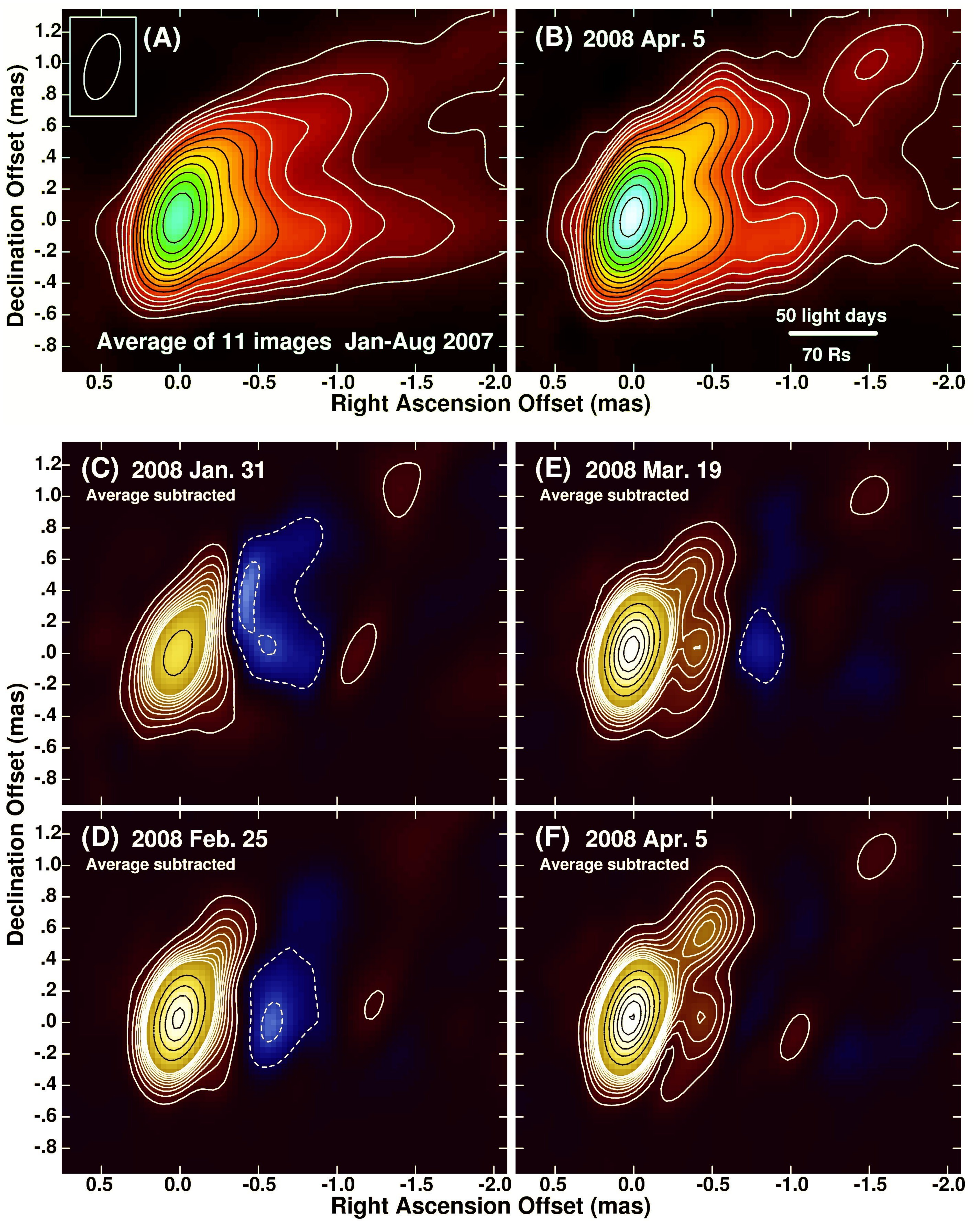,width=0.75\textwidth}
}

\caption{\label{fig3} \footnotesize VLBA images of M\,87 at $43 \,
\rm{GHz}$. ({\bf A}):  Average (hence smoothed) of 11 images from data
taken between January and August, 2007, well before the VHE and radio
flare. The contour levels start at $5$, $10$, $14.3$, and $20 \, \rm{mJy}$
per beam and increase from there by factors of $\sqrt{2}$. The restoring
beam used for all of the images is $0.21 \times 0.43 \, \rm{mas}$ ($30
\times 60 \, R_{\rm{s}}$) elongated in position angle $-16 \deg$, as shown
by the ellipse in the upper left corner. ({\bf B}): Image from 5 April
2008, with the same contours and colors as in (A). The linear scale in
light days and Schwarzschild radii is also shown. ({\bf C - F}):
Difference images for observations during the period of the radio flare
showing its effects. These were made by subtracting the average image (A)
from the individual epoch images.  The contours are linear with 10 (white)
at intervals of $7 \, \rm{mJy}$ per beam followed by the rest (black) at
intervals of $70 \, \rm{mJy}$ per beam; negative contours are indicated by
dashed lines. At the time of the VHE flare, the core flux density was
already above the average but the region of the jet between $-0.5$ and
$-1.0 \, \rm{mas}$ RA offset was below average, suggesting that there had
been a period of below-normal activity leading up to the flare and that
the radio flare may have begun before the VHE flare. The sequence shows
the substantial rise in the core flux density and the appearance of
enhanced emission along the inner jet. Image from \cite{JointPaper}.}

\end{figure}

Over at least the following two months, until the VLBA monitoring project
ended, the $43 \, \rm{GHz}$ radio flux density from the region within $1.2
\, \rm{mas}$ of the core rose by $30 \%$ as compared with its level at the
time of the start of the VHE flare and by $57 \%$ as compared with the
average level in 2007 (Fig.~\ref{fig2}). The resolution of the $43 \,
\rm{GHz}$ images corresponds to $30 \times 60 \, R_{\rm{s}}$ and the
initial radio flux density increase was located in the unresolved core.
The region around the core brightened as the flare progressed
(Fig.~\ref{fig3}), suggesting that new components were emerging from the
core. At the end of the observations, the brightened region extended about
$0.77 \, \rm{mas}$ from the peak of the core, implying an average apparent
velocity of $1.1 \, c$ ($c$ is the speed of light), well under the
approximately $2.3 \, c$ seen just beyond that distance in the first half
of 2007. Astrometric results obtained as part of the VLBA monitoring
program show that the position of the M\,87 radio peak, relative to M\,84,
did not move by more than $\sim$6~$R_{\rm{s}}$ during the flare,
suggesting that the peak emission corresponds to the nucleus of M\,87.

Because VHE, X-ray and radio flares of the observed magnitude are
uncommon, the fact that they happen together (chance probability of $P <
0.5 \%$, supporting online text) is good evidence that they are connected.
This is supported by our joint modeling of the VHE and radio light curves:
The observed pattern can be explained by an event in the central region
causing the VHE flare. The plasma travels down the jet and the effect of
synchrotron self-absorption causes a delay of the observed peak in radio
emission because the region is not transparent at radio energies at the
beginning of the injection (supporting online text, Sec.~3). The VLBI
structure of the flare along with the timing of the VHE activity, imply
that the VHE emission occurred in a region that is small when compared
with the VLBA resolution. Unless a source of infrared radiation is located
very close to the central black hole, which is not supported by current
observations \cite{MidIR_Emission}, TeV $\gamma$-ray photons can escape
the central region of M\,87 without being heavily absorbed through
$\rm{e}^{+}\rm{e}^{-}$ pair production \cite{ModelNeronov,ModelRieger}.

The light curve might indicate a rise in radio flux above the range of
variations observed in the past, starting before the first VHE flare was
detected. This could imply that the radio emission is coming from portions
of the jet launched from further out in the accretion disk than that
responsible for the VHE emission. However, it is difficult to derive a
quantitive statement on this, because no VHE data were taken in the week
previous to the flaring. Thus, an earlier start of the VHE activity cannot
be excluded, either.

A possible injection of plasma at the base of the jet observed at
optical and X-ray energies with a delayed passage through the radio core
$\sim$$10^{4} \, R_{\rm{s}}$ further down the jet~-- interpreted as a
standing shock and accompanied by an increase in radio emission~-- has
been discussed in the case of BL\,Lac \cite{Marscher} (with evidence for
VHE emission, see supporting online text for more details). M\,87 is much
closer than BL\,Lac and has a much more massive black hole, allowing the
VLBA to start resolving the jet collimation region whose size, from
general relativistic magnetohydrodynamic simulations
\cite{MD_Simulations}, is thought to extend over $\sim$1000~$R_{\rm{s}}$.
In case of M\,87 the radio core does not appear to be offset by more than
the VLBA resolution of $\sim$50~$R_{\rm{s}}$ from the black hole (see
supporting online text) and the jet has a larger angle to the
line-of-sight than in BL\,Lac. Thus the coincidence of the VHE and radio
flares (separated in photon frequency by 16 orders of magnitude),
constrains the VHE emission to occur well within the jet collimation
region.

\vspace*{0.5cm}

{\footnotesize {\bf Acknowledgements:} {\it H.E.S.S.:} The support of
the Namibian authorities and of the University of Namibia in
facilitating the construction and operation of H.E.S.S. is gratefully
acknowledged, as is the support by the German Ministry for Education and
Research (BMBF), the Max Planck Society, the French Ministry for
Research, the CNRS-IN2P3 and the Astroparticle Interdisciplinary
Programme of the CNRS, the U.K.  Science and Technology Facilities
Council (STFC), the IPNP of the Charles University, the Polish Ministry
of Science and Higher Education, the South African Department of Science
and Technology and National Research Foundation, and by the University
of Namibia. We appreciate the excellent work of the technical support
staff in Berlin, Durham, Hamburg, Heidelberg, Palaiseau, Paris, Saclay,
and in Namibia in the construction and operation of the equipment. {\it
MAGIC:} The collaboration thanks the Instituto de Astrof\'{\i}sica de
Canarias for the excellent working conditions at the Observatorio del
Roque de los Muchachos in La Palma, as well as the German BMBF and MPG,
the Italian INFN and Spanish MICINN.  This work was also supported by
ETH Research Grant TH 34/043, by the Polish MniSzW Grant N N203 390834,
and by the YIP of the Helmholtz Gemeinschaft. {\it VERITAS:} This
research is supported by grants from the U.S. Department of Energy, the
U.S. National Science Foundation and the Smithsonian Institution, by
NSERC in Canada, by Science Foundation Ireland and by the STFC in the
U.K. We acknowledge the excellent work of the technical support staff at
the FLWO and the collaborating institutions in the construction and
operation of the instrument. {\it VLBA:} The Very Long Baseline Array is
operated by the National Radio Astronomy Observatory, a facility of the
U.S. National Science Foundation, operated under cooperative agreement
by Associated Universities, Inc.}

\newpage
\section*{Supporting Online Material}



\subsection*{1) The geometry of the M\,87 jet}

\paragraph{General assumptions.} The luminosity distance of M\,87 is $D =
16.7 \, \rm{Mpc}$ \cite{Supp_M87_Distance}, so that an angle of $1$
milli-arc sec (mas) corresponds to $0.081 \, \rm{pc} = 2.5 \times 10^{17}
\, \rm{cm}$. The mass of the central black hole has been recently
determined using detailed modeling of long slit spectra of the central
regions of M\,87 to be $(6.0 \pm 0.5) \times 10^{9} \, \rm{M}_{\odot}$
\cite{BH_M87}, corrected for the distance we use\footnote{Gebhardt \&
Thomas (2009) assumed a distance of $17.9 \, \rm{Mpc}$. As those authors
point out, that mass is about $70\%$ higher than previous determinations
\cite{Supp_BH_M87_Old}, corrected for the different assumed distances. The
reason for such a large change is not fully explained yet. The use of the
new mass \cite{BH_M87} implies a higher resolution of our VLBA
observations in terms of $R_{\rm{s}}$, but does not substantially affect
our conclusions.}. The Schwarzschild radius is $R_{\rm{s}} \approx 1.8
\times 10^{15} \, \rm{cm}$; $1 \, \rm{mas}$ therefore corresponds to
approximately $140 \, R_{\rm{s}}$. The observed timing delay $\Delta
t_{\rm{obs}}$ between the onset of the VHE flare and the radio peak is
about $50 \, \rm{days}$ ($\sim 4 \times 10^{6} \, \rm{s}$). The following
equations hold: The observed velocity in units of the speed of light $c$
is calculated as $\beta_{\rm{obs}} = [\beta_{\rm{int}} \sin \theta] / [1 -
\beta_{\rm{int}} \cos \theta]$, the Doppler boost factor is $\delta = [
\Gamma (1 - \beta_{\rm{int}} \cos \theta) ]^{-1}$, the observed time delay
is $\Delta t_{\rm{obs}} = \Delta t_{\rm{int}} [ 1 - \beta_{\rm{int}} \cos
\theta) ]$, and the distance along the jet is $r_{\rm{int}} = r_{\rm{obs}}
/ \sin \theta \approx 3 \, r_{\rm{obs}}$ for $\theta = 20 \deg$ (see
below).

\paragraph{The jet orientation angle.} The most rapid optical proper
motions are observed along the jet with superluminal speeds
\cite{HST_Superluminal} of $\beta_{\rm{obs}} \approx 6$ (HST-1),
$\beta_{\rm{obs}} \approx 5$ (knot~D), and $\beta_{\rm{obs}} \approx 4$
(knot~E). On the other hand, the most rapid radio proper motions seen
from features moving through HST-1 indicate superluminal speeds in the
range of $\beta_{\rm{obs}} \approx 3-4$ \cite{Jet_And_TeV} and
through knot~D indicate a superluminal speed of $\beta_{\rm{obs}}
\approx 2.5$ \cite{Supp_ProperMotionInJet}. Assuming the jet plasma
moves at the speed of light ($\beta_{\rm{int}} = 1$), one obtains a
maximum viewing angle of $\cos (\theta_{\rm{max}}) =
[\beta_{\rm{obs}}^{2} - 1] / [\beta_{\rm{obs}}^{2} + 1 ]$. The highest
optically determined superluminal speed requires a jet viewing angle of
$\theta < 18.9 \deg$. On the other hand, the highest radio determined
superluminal speed requires $\theta < 28.0-36.9 \deg$ for
$\beta_{\rm{obs}} \approx 4-3$, respectively. Jet angles of $\theta =
30-45 \deg$ were derived \cite{VLBA_M87} based on radio
observations at $43 \, \rm{GHz}$, assuming the component to the East of
the core is a counter-jet, and with a velocity measurement based on only
one pair of observations. In this paper we assume a likely range of the
jet angle of $\theta =15 - 25 \deg$.

\paragraph{The jet opening angle.} The observations indicate the
following jet full opening angles as a function of the distance to the
core: $\psi_{\rm{obs}} (< 0.5 \, \rm{mas}) \approx 60 \deg$
\cite{JetFormation}, $\psi_{\rm{obs}} (1 - 5 \, \rm{mas}) \approx
12 \deg$ \cite{ImagingJetBase}, and $\psi_{\rm{obs}} (> 10 \,
\rm{mas}) \approx 6 \deg$ \cite{InnerJet}.  Assuming a jet viewing
angle of $\theta = 20 \deg$ the corresponding intrinsic full opening
angles are: $\psi_{\rm{int}} (< 0.5 \, \rm{mas}) \approx 30 \deg$,
$\psi_{\rm{int}} (1 - 5 \, \rm{mas}) \approx 4 \deg$, and
$\psi_{\rm{int}} (> 10 \, \rm{mas}) \approx 2 \deg$. The observed
pattern of opening angles suggest that the radio core corresponds to the
formation region of the jet.

\paragraph{The counter-jet.} The optical identification of an emission
feature observed at radio frequencies located $24 \, \rm{arc \, sec}$ away
from the nucleus in the direction opposite to the jet resulted in a first
indication of a counter-jet in M\,87 \cite{Supp_CounterJet}. This feature
is also seen in observations at mid-infrared frequencies
\cite{MidIR_Emission}.  Radio observations at wavelengths of $2 \,
\rm{cm}$ show clear indications of a counter-feature extending up to $3 \,
\rm{mas}$ from the radio core in the direction opposite to the jet
\cite{InnerJet}. While individual features move along the jet
(determined from 21 images taken over a time span of more than 10 years),
the counter feature seems to move in the opposite direction with an
apparent velocity of $(0.010 \pm 0.001) \, c$, strengthening the
counter-jet interpretation.  However, it cannot be fully excluded that
this apparent movement is a result of temporal under-sampling in the
kinematic analysis, although the individual tracked jet features give
consistent brightness temperatures across the observation epochs, which
would be unlikely in case of incorrect cross-identifications of the
components. The jet to counter-jet brightness ratio ($0.5 - 3.1 \,
\rm{mas}$) is $10-15$. A counter-feature is also observed at $22$ and $43
\, \rm{GHz}$, extending roughly $1 \, \rm{mas}$ from the core in the
direction opposite to the jet \cite{VLBA_M87}, see also
Fig.~\ref{fig1}D.  The jet to counter-jet brightness ratio is calculated
to be $\sim 14$. Marginal indication for an apparent velocity of $0.17
\,c$ of the counter-jet away from the core is found from three $43 \,
\rm{GHz}$ radio images. The observations, however, suggest a temporal
under-sampling of those data.

\paragraph{Position of the Black hole.} At radio frequencies of $86 \,
\rm{GHz}$, the core is no larger than $25 \times 7 \, R_{\rm{s}}$
\cite{Supp_MAS_Imaging_M87}, and its jet morphology is consistent with a
wide-opening angle jet base, as seen at $43 \, \rm{GHz}$, converging
near the point of the black hole. Farther out between about $1$ and $10
\, \rm{mas}$ the bright edges of the jet converge toward a point close
to the maximum extent of the counter-feature. Interpreting it as a part
of the jet would require that the jet before the radio peak is better
collimated than afterwards, although it appears to be as wide as the jet
itself. The astrometric results relative to M\,84 show that the root
mean square scatter of the position of the radio core is about $5
\times 2 \, R_{\rm{s}}$ along and across the beam with no clear
systematic motion (Davies et al., in preparation). If the radio core is
farther down the jet and the jet power is going up fractionally like the
flux density, this would imply a very stable position of the shock
region. The spatial stability is a reasonable assumption for the radio
core being located close to the black hole. Theoretical modelling also
supports the hypothesis that the $43 \, \rm{GHz}$ radio peak emission
results from the position of the black hole \cite{Supp_BH_Silouette}.

\subsection*{2) The VHE and radio observations and frequency of flares}

\paragraph{VHE.} The H.E.S.S., MAGIC and VERITAS collaborations operate
imaging atmospheric Che-renkov telescopes (IACTs) located in Namibia,
the Canary Islands (Spain) and Arizona (USA), respectively. The
telescopes measure cosmic $\gamma$-ray photons (entering the atmosphere
of the Earth) in an energy range of $0.1 \, \rm{TeV}$ up to several 10's
of TeV. M\,87 has been observed at those energies for the last ten
years. Except for the 2008 observation campaign, the observations were
scheduled in advance and did not follow any external or internal
triggers, leading to arbitrarily sampled light curves. During the
observations of the past 10 years only two episodes of flaring activity
have been measured: in 2005 \cite{M87_HESS} and in 2008 (reported
in this paper). For the first time, M\,87 was observed by H.E.S.S.,
MAGIC and VERITAS in a joint campaign for more than $120 \, \rm{h}$ in
2008 (more than $95 \, \rm{h}$ of data after quality selection). The
integral fluxes presented in this paper (Fig.~\ref{fig2}) were
calculated\footnote{the H.E.S.S. flux points~-- measured with a higher
energy threshold~-- were extrapolated down from $\sim 1 \, \rm{TeV}$.}
under the assumption that the spectrum of M\,87 is described by a
power-law function $\rm{d}N / \rm{d}E \propto E^{-2.3}$
\cite{M87_MAGIC}. Any correlation between the spectral shape and
the flux level has not yet been established for M\,87. The relative
frequency of flaring activity was estimated by fitting the
night-by-night binned light curves as measured by H.E.S.S., MAGIC and
VERITAS with a constant function (using all available data~-- partly
archival~-- from 2004 to 2008). Subsequently flux nights with the most
significant deviation from the average were removed until the fit
resulted in a reduced $\chi^{2}$ per degree of freedom of less than $1$;
all removed points corresponded to flux values higher than the average.
The light curves are compatible with constant emission for 49 out of 53
nights (H.E.S.S., 2004-2008), 12 out of 21 nights (MAGIC, 2008) and 50
out of 51 nights (VERITAS, 2007-2008). Combining these numbers one finds
flaring activity in the so far recorded data in 14 out of 125 nights of
observations, resulting in a relative frequency of flares on the order
of $10 \%$ of all observed nights.  Almost all data were recorded
arbitrarily and except for four nights (with a time difference of $\sim
0.5 \, \rm{days}$ between the VERITAS and the H.E.S.S./MAGIC
observations) all observations were separated in time by more than one
day. Therefore we assume that this number gives an estimate of the
general chance to measuring a VHE $\gamma$-ray flare from M\,87.
However, the relative frequency of flaring activity is overestimated by
the fact that the 2008 observations were intensified for some nights
during the high flux state following the VHE trigger by MAGIC
\cite{M87_MAGIC}.

\paragraph{X-ray.} M\,87 was regularly observed at X-ray energies with
Chandra, resulting in 61 measurements of the X-ray flux of the nucleus
during the last ten years \cite{TimeScalesHST}. Three measurements
exceed a flux level higher than 2 times the root mean square (RMS) of
the average flux of all data points (relative occurrence of $\sim 5
\%$). Only one measurement exceeds the level of $3 \, \rm{RMS}$ which
was taken during the radio flare with a deviation of $\sim 4.3 \,
\rm{RMS}$ (Fig.~\ref{fig2} in the main text).

\paragraph{Radio.} Throughout 2007, M\,87 was observed with the VLBA on
a regular basis roughly every three weeks \cite{Supp_VLBA_MovieM87}. The
aim of this 'movie project' was to study morphological changes of the
plasma jet with time. Preliminary analysis of the first 7 months showed
a fast evolving structure, somewhat reminiscent of a smoke plume, with
apparent velocities of about twice the speed of light. These motions
were faster than expected so the movie project was extended from January
to April 2008 with a sampling interval of 5 days. A full analysis of
these data is in progress and details will be published elsewhere. The
observed radio flux densities reached at the end of the 2008
observations, roughly 2 months after the VHE flare occurred, are larger
than seen in any previous VLBI observations of M\,87 at this frequency,
including during the preceding 12 months of intensive monitoring, in 6
observations in 2006 and in individual observations in 1999, 2000, 2001,
2002, and 2004 \cite{VLBA_M87}. The MOJAVE project web
site\footnote{http://www.physics.purdue.edu/MOJAVE/} gives $15 \,
\rm{GHz}$ VLBA flux densities at 27 epochs since 1995, with the highest
flux value measured on May 1, 2008\footnote{This is only a few weeks
later than the maximum of the $43 \, \rm{GHz}$ flux reported in this
paper. However, no further $15 \, \rm{GHz}$ data for 2008 are listed on
the MOJAVE web page.}; most of the data (except the last two epochs) are
published in \cite{MOJAVE_Fluxes}. Assuming a flare duration of
$\sim$4~months, a similar flare was not observed during a total period
of $5 \times 4 = 20$ months based on the 5 observations from 2004 and
earlier, which are well separated, 8 months taking into account overlap
based on the 2006 pilot observations spread over 4 months, and 14 months
during the 2007/2008 monitoring including 2 months before the start but
not including the time during the observed flare. That is 42 months
total for which a similar flare was not in progress.  By the same
accounting, there are 4 months with a flare.  So the probability of a
radio flare being in progress at any given time is $4/46 \approx 10\%$,
suggesting that radio flares of the observed magnitude are uncommon.

\paragraph{The observed radio/X-ray/VHE-pattern.} The probability of
observing $k$ out of $n$ nights exceeding a flux baseline (for which the
chance probability is $p$) is: $p(n,k) = {n \choose k} \cdot p^{k} \cdot
(1-p)^{n-k}$. The probability of observing $k$ or more flare nights is
$P(n, \geq k) = \sum_{j=k}^{n} p(n,j)$. A joint estimated of the chance
probability of the observed pattern is difficult, since the
characteristic time scales of flux changes are different and the data
are not sampled equally, or are partly sampled with higher frequencies
as the characteristic time scale of the flux changes (over-sampling) or
much less frequent (under-sampling). However, if defining a time window
covering the whole increase of the radio flux between end of January to
mid of April one finds 8 out of 40 VHE measurements ($P_{\rm{VHE}}$) and
1 out of 2 X-ray measurements ($P_{\rm{X-ray}}$) exceeded their
baselines.  The chance probability of observing this pattern during the
radio flare is on the order of $P = P_{\rm{VHE}} \cdot P_{\rm{X-ray}} =
0.026 \cdot 0.095 < 0.5 \%$. The fluxes in the radio, X-ray, and VHE
bands reached their highest archival level during the defined time
window.

\subsection*{3) Time-dependent modelling of the radio emission as
synchrotron emission from a slow outer sheath}

\paragraph{The model.} Whereas the VHE $\gamma$-ray emission from M\,87
varies on time scales of a few days, the $43 \, \rm{GHz}$ radio emission
from the nucleus steadily increases over a time period of two months
(Fig.~\ref{fig2}). A model calculation was performed to test if the slow
variations of the radio flux can be explained with a self-absorbed
synchrotron model of electrons injected into a "slow outer sheath" of jet
plasma. The slow outer sheath has the geometry of a hollow cone, an
assumption which is supported by the edge-brightened structure of the jet
observed at radio frequencies (Fig.~\ref{fig1}D). As the plasma travels
down the jet, it expands, leading to a decline of the frozen-in magnetic
field and to adiabatic cooling of the electrons. In the model, a
$\gamma$-ray flare leads to the injection of radio-emitting plasma at the
base of the jet and at the base of the slow outer sheath. The model
assumes that the VHE $\gamma$-rays are produced very close to the black
hole, e.g. in the black hole magnetosphere and does not attempt to
describe the production mechanism. It merely uses the observed
$\gamma$-ray fluxes to normalize the energy spectrum of the electrons
responsible for the radio emission. In the beginning, the radio-emitting
plasma is optically thick, and the synchrotron emission cannot escape.
Owing to the adiabatic expansion, the plasma eventually becomes optically
thin leading to a radio flare. The radio flare dies down owing to the
decline of the magnetic field and the adiabatic cooling of the electrons.

Following the injection of radio-emitting plasma at time $t_{0}$, a ring
of plasma with radius $R = R_{\rm{s}} + \beta_{\rm{jet}} c \sin \alpha
(t-t_{0})$ (with the thickness of the radio bright sheath being 1/5th of
the cone radius) travels down the jet. The emission of the ring is
computed in the frame of the moving plasma, assuming that the magnetic
field scales as $B \propto 1/R$, and the electrons cool adiabatically.
The calculation uses the standard equations for Lorentz transforming the
emission of different sections of the ring, see for example
\cite{Supp_IC_Reflection}. Taking into account light travel time
effects, the received radio flux at $43 \, \rm{GHz}$ is computed. The
overall normalization of the electron energy spectrum is adjusted to
reproduce the observed radio flux. The results of the model strongly
depend on the choice of the minimal radius $R_{\rm{min}} = R_{\rm{s}}$,
and weakly depend on the choices of the magnetic field $B$, the jet
opening angle $\alpha$, and and the thickness of the sheath.

\begin{figure}[t]

\begin{minipage}[c]{0.64\textwidth}
\epsfig{file=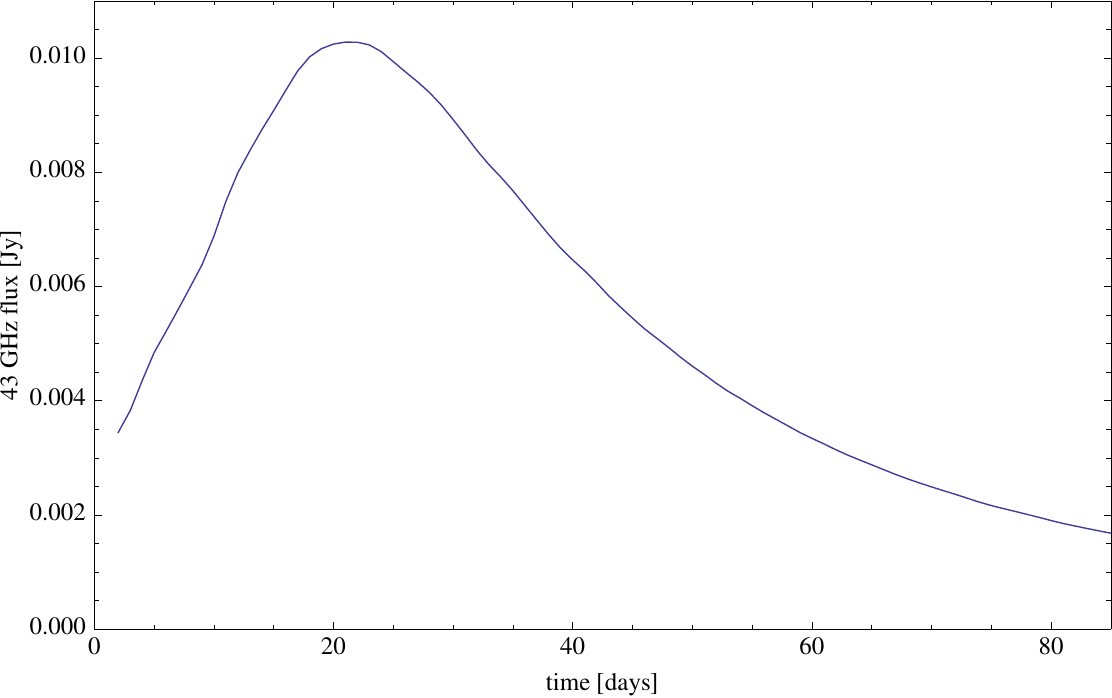,width=0.98\textwidth}
\end{minipage}
\begin{minipage}[c]{0.35\textwidth}

\caption{\label{figS1} \footnotesize Simulated $43 \, \rm{GHz}$ light
curve resulting from a single injection of radio-emitting plasma at time
$t = 0$ ($\Gamma = 1.01$, $\beta_{\rm{jet}} = 0.14$, $B = 0.5 \, \rm{G}$).
Initially, the plasma is opaque owing to synchrotron self-absorption. As
the plasma travels away from the point of injection, it expands and
becomes transparent. The expansion leads to a decrease of the magnetic
field and to adiabatic cooling of the electrons, and thus to the decline
of the radio emission. The small deviations during the injection stage are
due to numerical effects.}

\end{minipage}

\end{figure}

\begin{figure}[t]

\begin{minipage}[c]{0.64\textwidth}
\epsfig{file=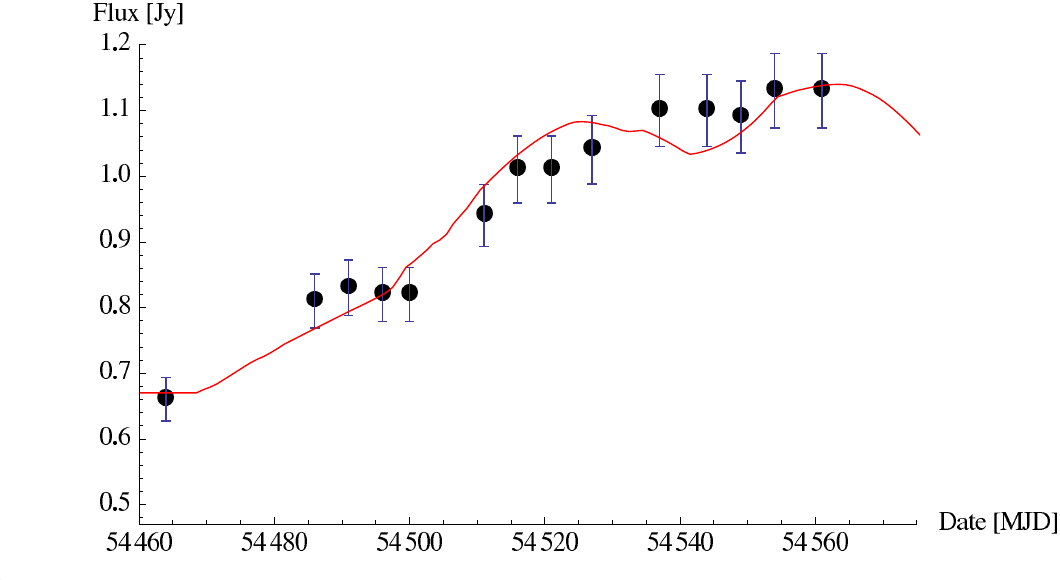,width=0.98\textwidth}
\end{minipage}
\begin{minipage}[c]{0.35\textwidth}

\caption{\label{figS2} \footnotesize Observed (data points) and modelled
(line) $43 \, \rm{GHz}$ radio flux. The modelled curve was obtained by
using the measured VHE $\gamma$-ray light curve as source function for
injecting radio-emitting plasma into the slow sheath of the jet. The Bulk
Lorentz factor used is $\Gamma = 1.01$ ($\beta_{\rm{jet}} =0.14$). No VHE
$\gamma$-ray observations were possible during phases of full moon,
leaving some uncertainties for the injection function.}

\end{minipage}

\end{figure}

\paragraph{Results.} We assume an intrinsic cone opening angle of $\alpha
= 5 \deg$, a jet angle of $\theta = 20 \deg$ and a rather low magnetic
field of $B = 0.5 \, \rm{G}$ at the base of the jet for which radiative
cooling can be neglected. The simulated radio light curve produced by a
single injection of radio-emitting plasma is shown in Fig.~\ref{figS1} for
an assumed bulk Lorentz factor of $\Gamma = 1.01$ ($\beta_{\rm{jet}} =
0.14$, giving the best fit result). The radio flux needs approximately
20~days to reach its maximum. Figure~\ref{figS2} shows the corresponding
radio light curve obtained when choosing a time-dependent electron
injection function proportional to the measured VHE $\gamma$-ray fluxes,
starting with the VHE data taken in January 2008. The spatial extent of
the predicted radio source after 50~days is $\sim 3$ light days and
therefore still within the central resolution element of the VLBA
observations (Fig.~\ref{fig3}). The model was chosen to minimize the
number of free parameters and assumptions. Other dependencies could affect
the results as follows: (i) If the emitting plasma is more compact (and
the volume filling factor is $<1$), it stays synchrotron self-absorbed for
a longer time, increasing the time lag between the $\gamma$-ray flares and
the rise of the radio flux. (ii) The emission volume may expand slower
than proportional to $t^{2}$, as assumed in the model. Slower expansion
would slow down the time scale for the rise and decay of the radio flux.
(iii) A higher value of $\beta_{\rm{jet}}$ would result in a faster radio
flare. (iv) All non-thermal particles are injected into the jet right at
the base of the slow sheath. This model assumption may not be accurate.
Additional non-thermal particles may be accelerated further downstream
which would lead to a longer duration of the radio flare. (v) If the
magnetic field at the base of the jet is stronger, radiative cooling is
not negligible any longer and fitting the data would require an assumption
of continued acceleration as the outer shell flows down the jet. (vi) In a
turbulent jet flow, efficient stochastic re-acceleration may occur, which
could change the picture. However, the current model calculations show
that synchrotron self-absorption may play a role in explaining the
observed slower turn-on of the radio emission.

\subsection*{4) Investigation of VHE $\gamma$-ray models for the M\,87
VHE/radio flare}

The key questions for the understanding of the VHE $\gamma$-ray emission
measured from the radio galaxy M\,87 and from the more than 20 known VHE
$\gamma$-ray blazars are: (i) What is the underlying particle
distribution which is accelerated, (ii) what are the mechanisms to
generate the $\gamma$-rays, and (iii) where is the region of the
emission located. In the following paragraphs a selection of models
discussed for M\,87 in the literature is investigated with a focus on
the question whether they can explain the observed VHE/radio light
curves. Note, however, that some of the models have difficulties in
explaining the observed hard VHE $\gamma$-ray spectra
\cite{M87_HESS,M87_MAGIC}, which will however not be discussed
in more detail.

\subsubsection*{4.1) Black hole magnetosphere models}

\paragraph{Models of VHE emission from the black hole magnetosphere.}
The electromagnetic mechanism of extraction of the rotational energy of
black holes by the Blandford \& Znajek scenario
\cite{Supp_EM_Extract_Kerr} seems to be a viable mechanism for powering
the relativistic jets of AGN. Particles can be accelerated by the
electric field of vacuum gaps in the black hole magnetosphere
\cite{ModelNeronov} (the electric field component parallel to the
ordered magnetic field is not screened out) or due to centrifugal
acceleration in an active plasma-rich environment, where the parallel
electric field is screened \cite{ModelRieger}.  Synchrotron and
curvature radiation of the charged particles, and inverse Compton
scattering of thermal photons can produce VHE $\gamma$-ray photons
\cite{Supp_Parallel_E}. An important question in this scenario is if the
$\gamma$-rays can escape the central region or if they are absorbed
through pair creation processes with either photons from the accretion
disk \cite{Jet_And_TeV} or infrared photons emitted by a potential
dust torus for which no clear observational evidence is found so far
\cite{MidIR_Emission,NoIR_Torus_Historical}. If M\,87 harbours a
non-standard (advection-dominated) accretion disk, $\gamma$-rays could
escape without being absorbed \cite{ModelRieger}. An alternative
scenario could be that the primary photons create a pair cascade whose
leakage produces the observed $\gamma$-ray emission
\cite{Supp_SynchrotronPairCascade}. The delayed radio emission could be
explained by the effect of synchrotron self-absorption (see Sec.~3) or
the time needed to cool the electrons before they dominantly emit
synchrotron radiation in the radio regime.

\subsubsection*{4.2) Hadronic jet models}

\paragraph{The model of Reimer et al. (2004).} In this model a primary
relativistic electron population is injected together with high-energy
protons into a highly magnetized emission region
\cite{ModelReimer}. The VHE emission is dominated by either
$\mu^{\pm} / \pi^{\pm}$ synchrotron radiation or by proton synchrotron
radiation. The low-energy component is explained by the synchrotron
emission of the electron population. However, the radio flux is
underestimated by the (steady-state) model (explanations are discussed
in the paper) so that a discussion of the observed radio/VHE flare is
beyond the scope of this particular model.

\subsubsection*{4.3) Leptonic jet models}

\paragraph{The model of Lenain et al. (2008).} In this model the
high-energy emission region (X-rays up to VHE) consists of small blobs
($\sim 10^{14} \, \rm{cm}$) travelling through the extended jet and
radiating at distances just beyond the Alfven surface
\cite{ModelLenain}. The emission takes place in the broadened jet
formation region, in the innermost part of the jet (corresponding to the
central resolution element in the $43 \, \rm{GHz}$ radio map). This
multi-blob model is a two-flow model, where the fast, compact blobs
contribute to X-rays and $\gamma$-rays through the synchrotron
self-Compton mechanism, and are embedded in an extended, diluted and
slower jet emitting synchrotron radiation from radio to optical
frequencies. Even though this model describes only steady state
emission, the observed radio/VHE variability can be discussed
qualitatively. For instance, a sudden rise of the density of the
underlying leptonic population at the stationary shock (i.e. in the
blobs) translates into a flare of X-rays and $\gamma$-rays, but no
immediate rise of radio emission is expected, because the emission
volume is synchrotron self-absorbed at radio frequencies, see Sec.~3.
However, as the flare propagates into the extended, less magnetized,
neighbouring jet, the leptons in the jet are energized and could cool by
emitting at radio frequencies with some delay, creating a diluted radio
flare in response to the VHE flare. An alternative scenario by Giannios
et al. (2009) explains the fast variability of VHE $\gamma$-ray
radiation in blazars as a possible result from large Lorentz factor
(100) filaments within a more slowly moving jet flow
\cite{Supp_Giannios}.

\paragraph{The model of Tavecchio and Ghisellini (2008).} According to
this model \cite{ModelTav} the jet consists of a fast spine and
slow sheath layer. The photons from the fast spine are external-Compton
boosted to VHE by the slower sheath. In this framework one may assume
that the VHE flare comes from near to the core and the time lag to the
radio maximum is entirely a result of propagation of some disturbance
down the jet and the associated reduction in synchrotron
self-absorption. The X-rays should be synchrotron emission that is not
self-absorbed and flare at about the same time as the VHE $\gamma$-rays. 
In the radio data we see significant edge brightening suggestive of a
de-boosted spine. Note, however, that edge brightening like that
observed can also be produced by enhanced surface emissivity. In the
model the radio emission originates from a region different from that
producing the VHE emission, so that a strict flux correlation is not
required. Detailed modeling would be needed to explain the observed
light curves in the framework of this model.

\paragraph{The model of Georganopoulos et al. (2005).} In this model the
jet decelerates over a length of $0.1 \, \rm{pc} = 3 \cdot 10^{17} \,
\rm{cm}$ \cite{ModelGeorg}. The VHE emission is assumed to come from the
fast moving part near the jet base by inverse-Compton scattering of
low-frequency photons from the slower moving part of the jet. The model
calculations are steady state so that they are difficult to apply to an
ejection event. However, one can assume that the VHE flare is directly
associated with the ejection event. As the disturbance propagates down
the jet and decelerates we see the radio rise later as a result of
synchrotron self-absorption effects, similar to the model described in
Sec.~3. Here the X-ray flux can still rise and fall with the VHE if it
comes directly from the disturbance; the majority of the observed power
comes from the slower part of the flow, but this is a more or less
steady state result. A similar model by Levinson (2007) of radiative
decelerating blobs in the jets of VHE $\gamma$-ray emitting blazars is
described in \cite{Supp_ModelLevinson}.

\subsubsection*{4.4) Jet base / standing shock models}

\paragraph{The model of Marscher et al. (2008).} The model is based on
the observation of a double flare in BL\,Lac \cite{Marscher}. The
first flare is seen at X-ray/optical energies accompanied by
polarization measurements at a date of 2005.82. It indicates an
injection event into the jet acceleration and collimation region near to
the black hole. The flare was followed by the appearance of a new radio
component in VLBA images that approached and passed through the radio
core, accompanied by a second X-ray/optical flare at 2005.92, where the
$14.5 \, \rm{GHz}$ radio flux begins to increase and peaks at about
2006.0. This is interpreted as the passage of the disturbance through a
standing shock in the jet at a distance of $\sim 10^{4} \, R_{\rm{s}}$,
see Fig.~3 in \cite{Marscher}. VHE $\gamma$-ray emission has been
detected in 2005, however, no evidence for flux variability has been
found in the data. In this picture the radio core is located at the
standing shock and the radio emission in the acceleration and
collimation region may be (i) either intrinsically weak or (ii)
synchrotron self-absorbed. The non-coincidence between the radio peak
and the second peak in the X-ray/optical is proposed to arise from the
longer lifetime of particles radiating at radio frequencies. Since the
disturbance passes down the expanding jet the radio emission might last
longer than the second X-ray/optical flare but does not increase in
strength after the disturbance passes through the shock. 

Applying this model to M\,87 one can assume that the observed VHE flare
corresponds to (A) the first or (B) the second flare. A: The VHE flare
indicates the injection at the base of the jet and the increase of the
radio flux corresponds to the passage through a standing shock that is
located at the M\,87 radio core. In this case, the counter-feature would
have to be interpreted as the jet before the standing shock. Although
the peak of the radio emission is delayed, the radio flux started to
increase at about the same time as the VHE flare, which indicates that
the two emission regions are not spatially separated, making this
scenario unlikely. B: The VHE flare indicates the passage of the
disturbance through the standing shock accompanied by a slow increase of
the radio flux. In this interpretation the first flare related to the
injection at the base of the jet would have been missed completely in
any of the wavelengths. There would still be a problem with a non time
coincident maximum in the VHE and radio emission. The VHE to radio lag
could be explained by synchrotron self-absorption (see Sec.~3.), which
would seem to require a coincidental juxtaposition of the standing shock
with just the right radio optical depth and subsequent optical depth
decline down the expanding jet. An alternative scenario could be that
the shock-accelerated particles causing the VHE emission cool rapidly
until they later emit photons dominantly at radio frequencies.

BL\,Lac is $16$ times farther away than M\,87, and the jet angle is
considerably smaller ($\theta \approx 7 \deg$) as compared to M\,87. The
black hole in M\,87, on the other hand, is $\sim 30$ times more
massive than the one in BL\,Lac. Therefore, our data have a $\sim
16$ times higher spatial resolution\footnote{even though the observations
in \cite{Marscher} use an angular resolution twice as high as ours at
$43 \, \rm{GHz}$} and provide a more than two orders of magnitude more
detailed insight into the jet physics on gravitational scales: In case of
the M\,87 observations presented here, $1 \, \rm{mas}$ corresponds to
$140 \, R_{\rm{s}}$, whereas in the case of the BL\,Lac
observations, $1 \, \rm{mas}$ corresponds to $\sim 70,000 \, R_{\rm{s}}$.
Although the Marscher et al.  model makes use of the observed VHE emission
in BL\,Lac, there is no experimental evidence that would constrain the
spatial region of that emission. Our observations connect the VHE emission
with the radio emission from the nucleus in M\,87 and therefore contrain
the VHE emission region to lie within the collimation region of the jet,
at maximum a few hundred $R_{\rm{s}}$ away from the black hole.



\section*{Full list of authors}

{\bf The VERITAS Collaboration:} 
V.~A.~Acciari$^{1}$,
E.~Aliu$^{2}$,
T.~Arlen$^{3}$,
M.~Bautista$^{4}$,
M.~Beilicke$^{5}$,
W.~Benbow$^{1}$,
S.~M.~Bradbury$^{6}$,
J.~H.~Buckley$^{5}$,
V.~Bugaev$^{5}$,
Y.~Butt$^{7}$
% ...

\vspace*{0.4cm}

{\it $^{1}$Fred Lawrence Whipple Observatory, Harvard-Smithsonian Center for Astrophysics, Amado, AZ 85645, USA}, 
{\it $^{2}$Department of Physics and Astronomy and the Bartol Research Institute, University of Delaware, Newark, DE 19716, USA}, 
{\it $^{3}$Department of Physics and Astronomy, University of California, Los Angeles, CA 90095, USA}, 
{\it $^{4}$Physics Department, McGill University, Montreal, QC H3A 2T8, Canada}, 
{\it $^{5}$Department of Physics, Washington University, St. Louis, MO 63130, USA}, 
{\it $^{6}$School of Physics and Astronomy, University of Leeds, Leeds, LS2 9JT, UK}, 
{\it $^{7}$Harvard-Smithsonian Center for Astrophysics, 60 Garden Street, Cambridge, MA 02138, USA}, 
{\it $^{8}$Argonne National Laboratory, 9700 S. Cass Avenue, Argonne, IL 60439, USA}, 
{\it $^{9}$School of Physics, University College Dublin, Belfield, Dublin 4, Ireland}, 
{\it $^{10}$School of Physics, National University of Ireland, Galway, Ireland}, 
{\it $^{11}$Astronomy Department, Adler Planetarium and Astronomy Museum, Chicago, IL 60605, USA}, 
{\it $^{12}$Department of Physics, Purdue University, West Lafayette, IN 47907, USA }, 
{\it $^{13}$Department of Physics and Astronomy, Barnard College, Columbia University, NY 10027, USA}, 
{\it $^{14}$Santa Cruz Institute for Particle Physics and Department of Physics, University of California, Santa Cruz, CA 95064, USA}, 
{\it $^{15}$Laboratoire Leprince-Ringuet, Ecole Polytechnique, CNRS/IN2P3, F-91128 Palaiseau, France}, 
{\it $^{16}$Department of Physics and Astronomy, University of Utah, Salt Lake City, UT 84112, USA}, 
{\it $^{17}$Enrico Fermi Institute, University of Chicago, Chicago, IL 60637, USA}, 
{\it $^{18}$Department of Physics and Astronomy, Iowa State University, Ames, IA 50011, USA}, 
{\it $^{19}$Department of Physics and Astronomy, University of Iowa, Van Allen Hall, Iowa City, IA 52242, USA}, 
{\it $^{20}$Department of Physics, Pittsburg State University, 1701 South Broadway, Pittsburg, KS 66762, USA}, 
{\it $^{21}$Department of Physics, Anderson University, 1100 East 5th Street, Anderson, IN 46012}, 
{\it $^{22}$Department of Life and Physical Sciences, Galway-Mayo Institute of Technology, Dublin Road, Galway, Ireland}, 
{\it $^{23}$European Southern Observatory, Karl-Schwarzschild-Strasse 2, 85748 Garching, Germany}, 
{\it $^{24}$Kavli Institute for Cosmological Physics, University of Chicago, Chicago, IL 60637, USA}, 
{\it $^{25}$Department of Applied Physics and Instrumentation, Cork Institute of Technology, Bishopstown, Cork, Ireland}, 
{\it $^{26}$National Radio Astronomy Observatory, Socorro, NM 87801, USA}, 
{\it $^{27}$Physics Department, 333 Workman Center, New Mexico Institute of Mining and Technology, 801 Leroy Place, Socorro, NM 87801, USA}, 
{\it $^{28}$Department of Physics and Astronomy, University of Alabama, Tuscaloosa, AL 35487, USA}, 
{\it $^{29}$ISR-2, MS-D436, Los Alamos National Laboratory, Los Alamos, NM 87545, USA}, 
{\it $^{30}$Department of Astronomy, University of California, Los Angeles, CA 90095-1547, USA}, 
{\it $^{31}$Max-Planck-Institut f\"ur Kernphysik, P.O. Box 103980, D-69029 Heidelberg, Germany}, 
{\it $^{32}$Yerevan Physics Institute, 2 Alikhanian Brothers St., 375036 Yerevan, Armenia}, 
{\it $^{33}$Centre d'Etude Spatiale des Rayonnements, CNRS/UPS, 9 av. du Colonel Roche, BP 4346, F-31029 Toulouse Cedex 4, France}, 
{\it $^{34}$Universit\"at Hamburg, Institut f\"ur Experimentalphysik, Luruper Chaussee 149, D-22761 Hamburg, Germany}, 
{\it $^{35}$Institut f\"ur Physik, Humboldt-Universit\"at zu Berlin, Newtonstr. 15, D-12489 Berlin, Germany}, 
{\it $^{36}$LUTH, Observatoire de Paris, CNRS, Universit\'e Paris Diderot, 5 Place Jules Janssen, 92190 Meudon, France}, 
{\it $^{37}$IRFU/DSM/CEA, CE Saclay, F-91191 Gif-sur-Yvette, Cedex, France}, 
{\it $^{38}$University of Durham, Department of Physics, South Road, Durham DH1 3LE, U.K.}, 
{\it $^{39}$Unit for Space Physics, North-West University, Potchefstroom 2520, South Africa}, 
{\it $^{40}$Laboratoire Leprince-Ringuet, Ecole Polytechnique, CNRS/IN2P3, F-91128 Palaiseau, France}, 
{\it $^{41}$Laboratoire d'Annecy-le-Vieux de Physique des Particules, CNRS/IN2P3, 9 Chemin de Bellevue - BP 110 F-74941 Annecy-le-Vieux Cedex, France}, 
{\it $^{42}$Astroparticule et Cosmologie (APC), CNRS, Universite Paris 7 Denis Diderot, 10, rue Alice Domon et Leonie Duquet, F-75205 Paris Cedex 13, France; UMR 7164 (CNRS, Universit\'e Paris VII, CEA, Observatoire de Paris)}, 
{\it $^{43}$Dublin Institute for Advanced Studies, 5 Merrion Square, Dublin 2, Ireland}, 
{\it $^{44}$Landessternwarte, Universit\"at Heidelberg, K\"onigstuhl, D-69117 Heidelberg, Germany}, 
{\it $^{45}$Laboratoire de Physique Th\'eorique et Astroparticules, Universit\'e Montpellier 2, CNRS/IN2P3, CC 70, Place Eug\`ene Bataillon, F-34095 Montpellier Cedex 5, France}, 
{\it $^{46}$Universit\"at Erlangen-N\"urnberg, Physikalisches Institut, Erwin-Rommel-Str. 1,D-91058 Erlangen, Germany}, 
{\it $^{47}$Laboratoire d'Astrophysique de Grenoble, INSU/CNRS, Universit\'e Joseph Fourier, BP 53, F-38041 Grenoble Cedex 9, France }, 
{\it $^{48}$Institut f\"ur Astronomie und Astrophysik, Universit\"at T\"ubingen, Sand 1, D-72076 T\"ubingen, Germany}, 
{\it $^{49}$LPNHE, Universit\'e Pierre et Marie Curie Paris 6, Universit\'e Denis Diderot Paris 7, CNRS/IN2P3, 4 Place Jussieu, F-75252, Paris Cedex 5, France}, 
{\it $^{50}$Charles University, Faculty of Mathematics and Physics, Institute of Particle and Nuclear Physics, V Hole\v{s}ovi\v{c}k\'{a}ch 2, 180 00}, 
{\it $^{51}$Institut f\"ur Theoretische Physik, Lehrstuhl IV: Weltraum und Astrophysik, Ruhr-Universit\"at Bochum, D-44780 Bochum, Germany}, 
{\it $^{52}$University of Namibia, Private Bag 13301, Windhoek, Namibia}, 
{\it $^{53}$Obserwatorium Astronomiczne, Uniwersytet Jagiello{\'n}ski, ul. Orla 171, 30-244 Krak{\'o}w, Poland}, 
{\it $^{54}$Nicolaus Copernicus Astronomical Center, ul. Bartycka 18, 00-716 Warsaw, Poland}, 
{\it $^{55}$School of Physics \& Astronomy, University of Leeds, Leeds LS2 9JT, UK}, 
{\it $^{56}$School of Chemistry \& Physics, University of Adelaide, Adelaide 5005, Australia}, 
{\it $^{57}$Toru{\'n} Centre for Astronomy, Nicolaus Copernicus University, ul. Gagarina 11, 87-100 Toru{\'n}, Poland}, 
{\it $^{58}$Instytut Fizyki J\c{a}drowej PAN, ul. Radzikowskiego 152, 31-342 Krak{\'o}w, Poland}, 
{\it $^{59}$European Associated Laboratory for Gamma-Ray Astronomy, jointly supported by CNRS and MPG}, 
{\it $^{60}$Supported by CAPES Foundation, Ministry of Education of Brazil}, 
{\it $^{61}$ETH Zurich, CH-8093 Switzerland}, 
{\it $^{62}$INAF National Institute for Astrophysics, I-00136 Rome, Italy}, 
{\it $^{63}$Universidad Complutense, E-28040 Madrid, Spain}, 
{\it $^{64}$Technische Universit\"at Dortmund, D-44221 Dortmund, Germany}, 
{\it $^{65}$Universitat Aut\`onoma de Barcelona, E-08193 Bellaterra, Spain}, 
{\it $^{66}$Universit\`a di Padova and INFN, I-35131 Padova, Italy}, 
{\it $^{67}$Inst. de Astrof\'{\i}sica de Canarias, E-38200 La Laguna, Tenerife, Spain}, 
{\it $^{68}$University of \L\'od\'z, PL-90236 Lodz, Poland}, 
{\it $^{69}$Deutsches Elektronen-Synchrotron (DESY), D-15738 Zeuthen, Germany}, 
{\it $^{70}$Max-Planck-Institut f\"ur Physik, D-80805 M\"unchen, Germany}, 
{\it $^{71}$Universit\`a  di Siena, and INFN Pisa, I-53100 Siena, Italy}, 
{\it $^{72}$Universitat de Barcelona (ICC/IEEC), E-08028 Barcelona, Spain}, 
{\it $^{73}$Universit\"at W\"urzburg, D-97074 W\"urzburg, Germany}, 
{\it $^{74}$IFAE, Edifici Cn., Campus UAB, E-08193 Bellaterra, Spain}, 
{\it $^{75}$Depto. de Astrofisica, Universidad, E-38206 La Laguna, Tenerife, Spain}, 
{\it $^{76}$Universit\`a di Udine, and INFN Trieste, I-33100 Udine, Italy}, 
{\it $^{77}$Institut de Cienci\`es de l'Espai (IEEC-CSIC), E-08193 Bellaterra, Spain}, 
{\it $^{78}$Inst. de Astrof\'{\i}sica de Andalucia (CSIC), E-18080 Granada, Spain}, 
{\it $^{79}$University of California, Davis, CA-95616-8677, USA}, 
{\it $^{80}$Tuorla Observatory, Turku University, FI-21500 Piikki\"o, Finland}, 
{\it $^{81}$Inst. for Nucl. Research and Nucl. Energy, BG-1784 Sofia, Bulgaria}, 
{\it $^{82}$INAF/Osservatorio Astronomico and INFN, I-34143 Trieste, Italy}, 
{\it $^{83}$ICREA, E-08010 Barcelona, Spain}, 
{\it $^{84}$Universit\`a  di Pisa, and INFN Pisa, I-56126 Pisa, Italy}, 
{\it $^{85}$supported by INFN Padova}, 
{\it $^{86}$deceased}.

\end{document}